\documentclass[12pt]{article}
\usepackage{graphicx}
\usepackage{cite}
\usepackage{amssymb,amsmath}
\usepackage[para]{threeparttable}
\usepackage{multirow}
\begin{document}
\title{THE POSSIBILITY OF BLOCKING THE PROCESS OF DNA BASE PAIRS OPENING BY HYDROGEN PEROXIDE}%
\author{Oleksii Zdorevskyi, Sergey N. Volkov\\
Bogolyubov Institute for Theoretical Physics, NAS of
Ukraine,\\14-b Metrolohichna Str., Kiev 03143, Ukraine \\
snvolkov@bitp.kiev.ua }\maketitle
\setcounter{page}{1}%
\maketitle
\begin{abstract}
One of the most progressive methods of cancer treatment is ion beam therapy. The simulations of water radiolysis show that in the cell medium the most long-living species are hydrogen peroxide ($H_2O_2$) molecules. But up to the present time the role of $H_2O_2$ molecules in the deactivation of cancer cells has not been determined yet. To understand the possible role of $H_2O_2$ in ion beam therapy, the competitive interaction of $H_2O$ and $H_2O_2$ molecules with nucleic bases in a pair on the different stages of genetic information transfer is studied in the present work. Atom-atomic potential functions method is used for the calculations.  It is shown, that some configurations of A{\textperiodcentered}T and G{\textperiodcentered}C complementary pairs are stabilized much better by $H_2O_2$ molecule compared to water molecule. Formation of such interaction complexes can terminate the processes of DNA unzipping by enzymes and consequently block genetic information transfer processes in cancer cells during ion beam therapy treatment. Experimental verification method of hydrogen peroxide interaction with nucleic base pairs is proposed.

\end{abstract}

\section{Intorduction}

Ion beam therapy is one of the most progressive methods for cancer treatment. Nowadays ion cancer facilities are built
throughout the Europe where patients undergo treatment on special accelerators. In the basis of ion therapy lies the
so-called Bragg effect~\cite{bragg1904}, which means that the maximum amount of energy is transferred to the medium by heavy ions at the end of their track in a certain localized area of space. This effect makes ion therapy especially effective for the treatment of tumors that are deep enough in the body.

In radiotherapy it is considered that to destroy the cancer cells, it is necessary in some way to deactivate their DNA~\cite{solovyov2010}. However, the definite mechanisms of action of high-energy ions on DNA of cancer cells have not been determined yet~\cite{KramerDurante2010}.

It is known that DNA macromolecule is situated in a cell in water-ionic solution. This solution stabilizes the double
helix, determines its shape, and, accordingly, affects its functioning in a living cell. Due to water radiolysis, which
takes place during ion beam therapy treatment, this environment changes significantly. A large number of species occurs
in solution: free radicals, secondary electrons, ions, as well as molecular products such as $H_2O_2$ and $H_2$. Until recently, in the literature~\cite{dsbreaks1998,solovyov2010} the most attention was paid to the DNA strand breaks, which occur due to the action of secondary electrons and free radicals. However, it is known that there are DNA reparation mechanisms, which can eliminate DNA strand breaks~\cite{NobelPrize2015}.

Experimental studies as well as the Monte Carlo simulation of water radiolysis~\cite{Kreipl2008,Uehara2006,Durante2018} showed that on the biological time scales one of the most long-living species are hydrogen peroxide molecules. But their role in ion beam therapy has not been discussed in the previous works properly.

In our paper~\cite{pyatEPJ2015}, the new hypothesis was proposed. According to this hypothesis, a hydrogen peroxide molecule can form stable complexes with active DNA sites, thus blocking the processes of genetic information transfer in living cells. To prove the hypothesis, in our works the interaction of $H_2O_2$ molecules with nonspecific DNA recognition sites - phosphate groups ($PO_4$)~\cite{pyatEPJ2015,piatBiophysVisnyk}, as well as with specific DNA recognition sites - nucleic bases~\cite{zdorevskyi2018blocking} was investigated. Methods of atom-atom potential functions and density functional theory were used for the calculations. Results of both methods revealed that the hydrogen peroxide molecule can form a complex with DNA phosphate groups that is no less stable than the same complex with water molecule. Also the definite sites of Adenine (A), Thymine (T), Guanine (G) and Cytosine (C) nucleic bases, where the $H_2O_2$ molecule can interact much stronger than the water molecule, are determined. These interactions can block processes of DNA recognition by the enzyme.

It should be noted, that in the DNA double helix, the nucleic bases form complementary pairs - A{\textperiodcentered}T and G{\textperiodcentered}C. During genetic information transfer the complementary base pairs become to be open for the interaction with the surrounding molecules, i.e. expose their atomic groups to the solvent. The base pair opening can be blocked by $H_2O_2$ molecules during ion beam therapy treatment and can serve as an evidence of our proposed mechanism of DNA deactivation. It is sufficient that now base pairs opening can be investigated experimentally with the help of single-molecule manipulation technique~\cite{bustamante2003,bockelmann2002,ritort2010}, therefore can be used for the direct observation of the interaction of $H_2O_2$ molecules with DNA.

In Sec. \ref{methods} the methodology of our calculations is described. In Sec. \ref{results}, stability of the complexes consisting of {\textquoteleft}preopened{\textquoteright} and {\textquoteleft}stretched{\textquoteright} A{\textperiodcentered}T and G{\textperiodcentered}C base pairs together with hydrogen peroxide and water molecule is studied. The possibility of blocking the process of the nucleic base pairs opening by $H_2O_2$ molecules is considered.
In Sec \ref{experiment}, the experimental observation method of the interaction of  hydrogen peroxide molecules with DNA nucleic bases is discussed.

\section{Calculation methods}
\label{methods}

In the present work we will perform our calculations with the help of atom-atom potential function method. This method is now widely used in such force fields as CHARMM and
AMBER~\cite{Charmm,Amber,LaveryMolDynRev} for studying the structure of molecular complexes. Calculations are made using GNU Octave software package~\cite{octave}. In the framework of this
method, the energy of intermolecular interaction consists of van der Waals interactions, hydrogen bonds and Coulomb
interactions:

\begin{equation}
E\left(r\right)=\sum
_{i,j}(E_{\mathit{vdW}}\left(r_{\mathit{ij}}\right)+E_{\mathit{HB}}\left(r_{\mathit{ij}}\right)+E_{\mathit{Coul}}\left(r_{\mathit{ij}}\right))
\end{equation}

In the framework of the present method we consider all the covalent bonds and
angles as rigid.

Van der Waals's interaction is described by Lennard-Jones's {\textquoteleft}6-12{\textquoteright} potential:

\begin{equation}
E_{\mathit{vdW}}\left(r_{\mathit{ij}}\right)=-\frac{A_{\mathit{ij}}}{r_{\mathit{ij}}^6}+\frac{B_{\mathit{ij}}}{r_{\mathit{ij}}^{12}},
\end{equation}
where the parameters $A^{(10)}_{ij}$ , $B^{(10)}_{ij}$ , $A_{ij}$ , $B_{ij}$ are taken from the works~\cite{PoltevShul1986,Poltev1980}.

The energy of the hydrogen bond between atoms i and j is modeled by the modified Lenard-Jones potential {\textquoteleft}10 -12{\textquoteright}:

\begin{equation}
E_{\mathit{HB}}\left(r_{\mathit{ij}}\right)=\left[-\frac{A_{\mathit{ij}}^{\left(10\right)}}{r_{\mathit{ij}}^{10}}+\frac{B_{\mathit{ij}}^{\left(10\right)}}{r_{\mathit{ij}}^{12}}\right]\cos
\varphi ,
\end{equation}
where $r_{ij}$ - the distance between the atoms $i$ and $j$,  $\varphi $  - the angle of the hydrogen bond. For example, when the hydrogen bond is $O-H ... N$ , then  $\varphi $  is an angle between the lines of covalent bond ($O - H$) and the hydrogen bond ($H...N$).

Coulomb interaction is described by the electrostatic potential:

\begin{equation}
E_{\mathit{Coul}}\left(r_{\mathit{ij}}\right)=\frac 1{4\pi \varepsilon _0\varepsilon
\left(r_{\mathit{ij}}\right)}\frac{q_iq_j}{r_{\mathit{ij}}},
\end{equation}
where $q_i$ and $q_j$ are the charges of the atoms $i$ and $j$ located at a distance $r_{ij}$ , $\varepsilon $\textsubscript{0} is the vacuum permittivity, and $\varepsilon$(r) is the dielectric permittivity of the medium.

The charges $q_i$, $q_j$ for nucleic bases were taken from the works~\cite{PoltevShul1986,Poltev1980}. Charges of $H_2O$ and $H_2O_2$ molecules were calculated from the condition that the dipole moment of water molecule should be equal to $d_{H2O} = 1,86$ $D$~\cite{waterDipolMoment}, and of hydrogen peroxide molecule $d_{H2O2} = 2,10$ $D$~\cite{peroxideDipolMoment}. Hence, for the $H_2O$ molecule we obtain the charges $q_H = 0,33e$, $qO = -0,66e$, and, accordingly, for $H_2O_2$ $q_H = 0,41e$, $q_O = -0,41e$. The values of charges on the atoms of $H_2O_2$ molecule are in good agreement with charges obtained  by quantum-chemical calculations in the work~\cite{Moin2012}. Also the same charge values are used in recently developed force field for hydrogen peroxide molecule~\cite{peroxideForceField}.

Since DNA in the living cell is situated in a water-ion solution, the interacting atoms are screened by water molecules. This leads to a weakening of the Coulomb interaction. Thus, more effective accounting of Coulomb interactions can be
achieved using the dependence of the dielectric permittivity upon distance ($\varepsilon $(r)), developed by Hingerty \textit{et al. }~\cite{hingerty} in the explicit form:

\begin{equation}
\label{for:hing}
\epsilon \left(r\right)=78-77\left(r_p\right)^2\frac{e^{r_p}}{\left(e^{r_p}-1\right)^2},
\end{equation}
where $r_p=r/2.5$.

\section{Calculation results}
\label{results}
It is known that the nucleic bases in the complementary pair have many degrees of freedom, which are defined by the standard nomenclature~\cite{diekmann1989definitions}. Since the base pairs are situated in a double helix, their structure is stabilized by the stacking interaction between adjacent pairs. This essentially limits the degrees of freedom that remove the bases beyond the plane of the pair. Consequently, in this paper, only the degrees of freedom of the bases in the plane of the pair ({\textquoteleft}stretch{\textquoteright}, {\textquoteleft}opening{\textquoteright}, {\textquoteleft}shear{\textquoteright}) are considered. In addition, the degree of freedom {\textquoteleft}propeller twist{\textquoteright} was considered (Appendix, Fig. \ref{fig:pathways}), which is essential to be taken into account due to the spatial structure of $H_2O_2$ molecule.

\begin{figure}% figure* for wide figure, [h] [!] to change the placement
\centering
\vskip1mm
\includegraphics[width=17cm]{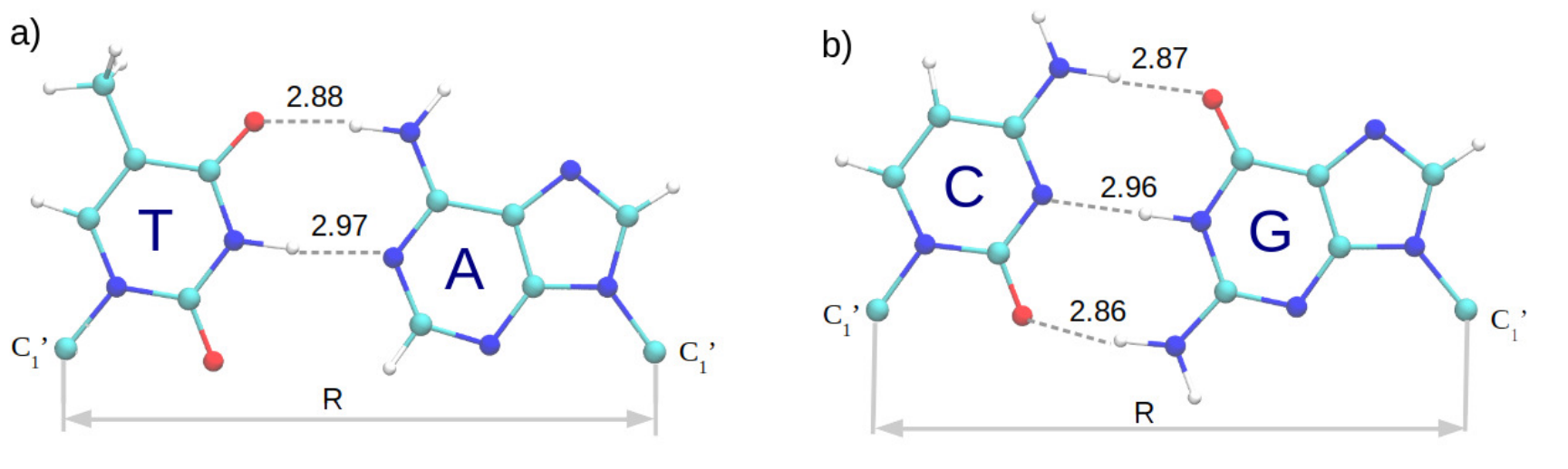}
\vskip-3mm\caption{ Watson-Crick configurations of the A{\textperiodcentered}T (a) and G{\textperiodcentered}C (b) complementary pairs. The numbers indicate the distance (in \AA) between the heavy atoms in the corresponding hydrogen bonds. $R$ denotes the distance between the $C_1'$ atoms. }
\label{fig:WC}
\end{figure}

Firstly, in order to verify the correctness of the parameters chosen for our calculations, the stable Watson-Crick configuration of the complementary pairs of A{\textperiodcentered}T and G{\textperiodcentered}C were calculated (Fig. \ref{fig:WC}). As can be seen from Tabl. \ref{tab:WCpapameters}, the spatial structures of these configurations are close to those obtained from the X-ray structural analysis and the experiment on nuclear magnetic resonance. The main differences occur in the parameter {\textquoteleft}propeller twist{\textquoteright} due to the fact that the experiment measures pairs which are not isolated, but which constitute a structure of a double helix. These parameters are calculated using the \textit{3DNA} software package~\cite{3dna}. Visualization is made by VMD~\cite{vmd}. Also Tabl. 1 shows that the spatial structure of the pairs does not differ much, depending on whether the formula (\ref{for:hing}) is used in the calculations. At the same time, the difference in the interaction energy for G{\textperiodcentered}C pair is significant, which is due to the anomalous contribution of the Coulomb interaction. More details of the necessity of using the dependency (\ref{for:hing}) are discussed in~\cite{ourWorkUnz}. It was shown in~\cite{zdorevskyi2018blocking} that the calculation results of the water-water and peroxide-water complexes with the use of the dependence (\ref{for:hing}) are much closer to the results of quantum-chemical calculations than without this dependence. Therefore, all further calculations in this paper will be carried out taking into account formula (\ref{for:hing}).

\begin{figure}% figure* for wide figure, [h] [!] to change the placement
\centering
\vskip1mm
\includegraphics[width=8cm]{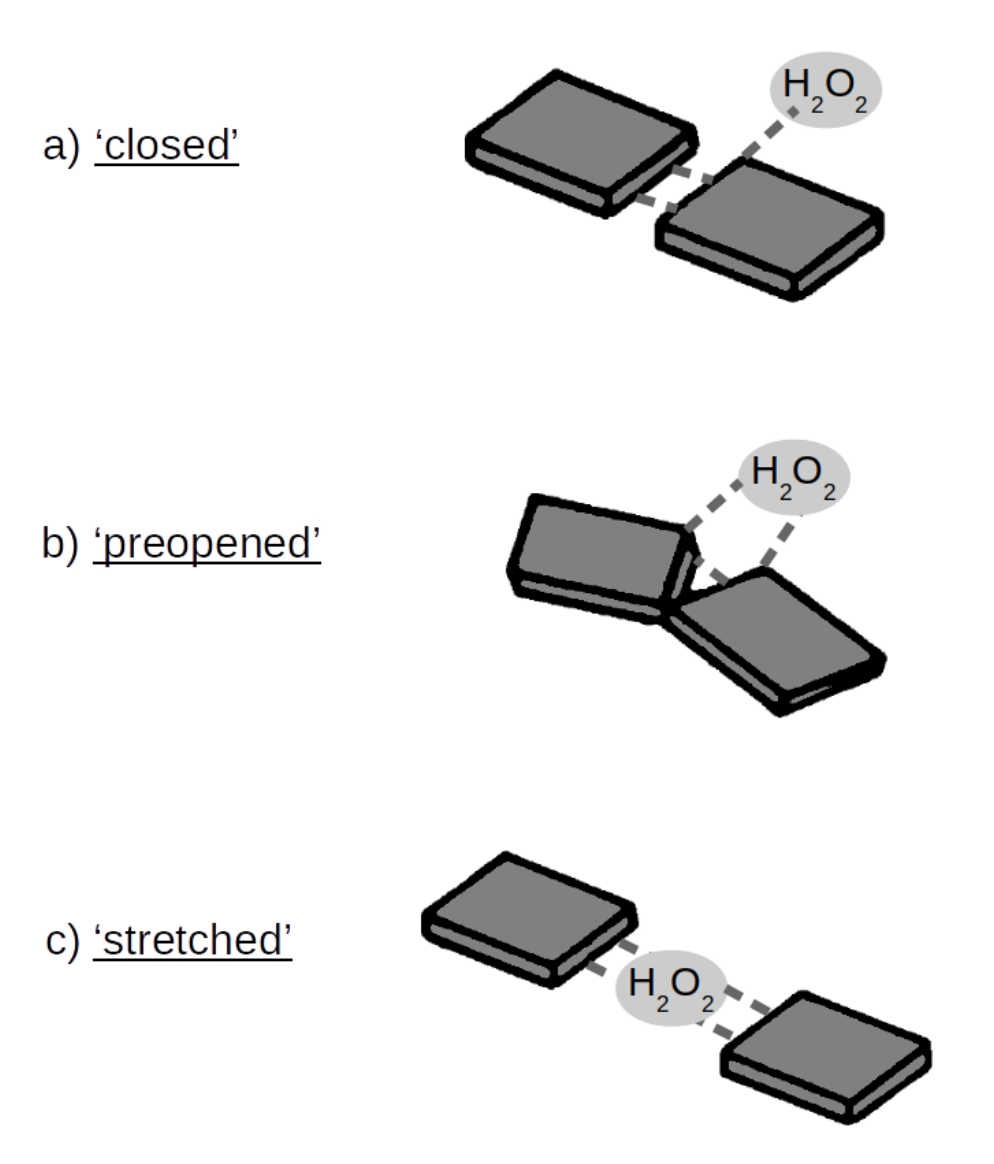}
\vskip-3mm\caption{ Complexes of nucleic base pairs with hydrogen peroxide molecules considered in the present work (and should occur on the pathway of base pairs opening during DNA unzipping): a){\textquoteleft}closed{\textquoteright} configuration; b) {\textquoteleft}preopened{\textquoteright} configuration; c) {\textquoteleft}stretched{\textquoteright} configuration.}
\label{fig:PairsConfigurations}
\end{figure}

\begin{table*}[t]

\noindent\caption{Structural parameters of Watson-Crick A{\textperiodcentered}T and G{\textperiodcentered}C base pairs according to the standard nomenclature (Appendix, Fig. \ref{fig:pathways}). The parameters {\textquoteleft}shear{\textquoteright}, {\textquoteleft}stretch{\textquoteright} are given in \textit{\AA};  {\textquoteleft}propeller twist{\textquoteright}, {\textquoteleft}opening{\textquoteright} in $degrees$.}\vskip3mm\tabcolsep4.5pt
\begin{threeparttable}[t]
\noindent{\footnotesize
\label{tab:WCpapameters}
\begin{tabular}{|c|cccc|ccccc|}
 \hline%
\multirow{3}{*}{Parameter} & \multicolumn{4}{c|}{\underline{A{\textperiodcentered}T}} & & \multicolumn{4}{c|}{\underline{G{\textperiodcentered}C}} \\

& \multirow{2}{*}{X-ray\tnote{1}} & \multirow{2}{*}{NMR\tnote{2}} & \multicolumn{2}{c|}{\underline{Our calculations}} & & \multirow{2}{*}{X-ray\tnote{1}} & \multirow{2}{*}{NMR\tnote{2}} & \multicolumn{2}{c|}{\underline{Our calculations}} \\
& & & vacuum & using (\ref{for:hing}) & & & & vacuum & using (\ref{for:hing}) \\
\hline%
 &  &  &  &  &  &  &  &  & \\
{\textquoteleft}\underline{stretch}{\textquoteright} & $-0.16 \pm 0.02$ & $-0.15 \pm 0.04$ & $-0.02$ & $-0.01$ & & $-0.24 \pm 0.02$ & $-0.3 \pm 0.03$ & $-0.14$ & $-0.14$   \\
 &  &  &  &  &  &  &  &  & \\
{\textquoteleft}\underline{shear}{\textquoteright} & $0.12 \pm 0.03$ & $0.00 \pm 0.04$ & $0.16$ & $0.18$ & & $-0.08 \pm 0.05$ & $0 \pm 0.07$ & $0.16$ & $0.23$ \\
 &  &  &  &  &  &  &  &  & \\
{\textquoteleft}\underline{opening}{\textquoteright} & $2.62 \pm 0.67$ & $-0.54 \pm 0.6$ & $-4.79$ & $-5.01$ & & $-2.13 \pm 0.4$ & $0.66 \pm 0.41$ & $-1.70$ & $-1.70$ \\
 &  &  &  &  &  &  &  &  & \\
{\textquoteleft}\underline{propeller}  & \multirow{2}{*}{$-16.95 \pm 0.34$} & \multirow{2}{*}{$-14.45 \pm 2.41$} & \multirow{2}{*}{$-2.46$} & \multirow{2}{*}{$3.58$} & & \multirow{2}{*}{$-8.15 \pm 1.49$} & \multirow{2}{*}{$-10.41 \pm 1.53$} & \multirow{2}{*}{$-0.53$} & \multirow{2}{*}{$-0.44$} \\
 \underline{twist}{\textquoteright}&  &  &  &  &  &  &  &  & \\
 &  &  &  &  &  &  &  &  & \\
  \hline%

\end{tabular}

\begin{tablenotes}\footnotesize
\item [1] calculated from the spatial structures of the Dickerson-Drew dodecamer (files 1bna.pdb, 7bna.pdb, 9bna.pdb, 436d.pdb) obtained by X-ray analysis. The values are averaged separately by the A{\textperiodcentered}T and the G{\textperiodcentered}C base pairs and the standard errors are calculated.

\item [2] calculated from the spatial structures of the Dickerson-Drew dodecamer (files 1duf.pdb, 1gip.pdb, 1naj.pdb, 2dau.pdb) obtained by the method of nuclear magnetic resonance. The values are averaged separately by the A{\textperiodcentered}T and the G{\textperiodcentered}C base pairs and the standard errors are calculated.
\end{tablenotes}

}
\end{threeparttable}

\end{table*}

The interaction of nucleic bases with water molecules was considered in a series of papers~\cite{Kryachko,Lavery2003,PoltevWater}. In the paper~\cite{gorb2004} hydration shells of A{\textperiodcentered}T and G{\textperiodcentered}C base pairs were calculated, that is,
simultaneous interaction with a large number of water molecules was considered. However, in the literature it was not
paid essential attention to the interaction of nucleic bases with hydrogen peroxide molecules (which, as described in
Sec. 1, appears in the cell as a result of ion beam treatment). Only the work~\cite{dobado1999} is known, where the
interaction of hydrogen peroxide molecules with the Adenine base was considered.

In the present paper calculations of interaction energy of complexes consisting of A{\textperiodcentered}T and G{\textperiodcentered}C nucleic base pairs with hydrogen peroxide and water molecules are carried out. In the work~\cite{volkovMolDynSym} it was shown that structures stabilized by water molecules occur on the pathway of base pairs opening during DNA unzipping. For the purposes of the present work, it is important to analyze the possibility of forming the simplest {\textquoteleft}non-Watson-Crick{\textquoteright} configurations of A{\textperiodcentered}T and G{\textperiodcentered}C base pairs, stabilized by hydrogen peroxide and water molecules, and to establish whether the formation of these complexes can block the genetic information transfer processes. Consequently, the complexes that include only one hydrogen peroxide or water molecule are taken into account.

In the present work we consider three configurations of nucleic base pairs with hydrogen peroxide molecules and water molecules. The complexes consisting of complementary A{\textperiodcentered}T and G{\textperiodcentered}C base pairs and the hydrogen peroxide or water molecule that interacts with the base from the side of the major groove we denote as {\textquoteleft}closed{\textquoteright} pairs (Fig. \ref{fig:PairsConfigurations} a). The configuration of the pair where the {\textquoteleft}opening{\textquoteright} pathway dominates, we denote {\textquoteleft}preopened{\textquoteright} (Fig. \ref{fig:PairsConfigurations} b), and those in which the {\textquoteleft}stretch{\textquoteright} pathway dominates will be denoted as {\textquoteleft}stretched{\textquoteright} pair (Fig. \ref{fig:PairsConfigurations} c).

Firstly, let us calculate {\textquoteleft}closed{\textquoteright} configurations. It can be seen, that the interaction with these molecules almost does not change the geometry of the Watson-Crick pairs (Tabl. \ref{tab:results}, Appendix, Fig. \ref{fig:Closed_H2O_H2O2}).

On Fig. \ref{fig:Preopened_H2O_H2O2} (Appendix) stable configurations of the {\textquoteleft}preopened{\textquoteright} A{\textperiodcentered}T and G{\textperiodcentered}C base pairs with water and hydrogen peroxide molecules are shown. For the convenience of the analysis, in Tabl. \ref{tab:results} only structural parameters $\Delta R=R-R_{WC}$ (difference in $C_1'C_1'$ distances (Fig. \ref{fig:WC} a) in the corresponding and Watson-Crick pair), {\textquoteleft}opening{\textquoteright} and {\textquoteleft}propeller{\textquoteright}, as well as interaction energies are shown. From Tabl. \ref{tab:results} it can be seen that the configurations of {\textquoteleft}preopened{\textquoteright} A{\textperiodcentered}T pairs with water and hydrogen peroxide molecules have almost identical parameters. The {\textquoteleft}opening{\textquoteright} parameter has the maximum value, the rest of the parameters do not change significantly. It also should be noted that the {\textquoteleft}propeller twist{\textquoteright} parameter almost does not take the bases out of the plane of the pair. Moreover, configurations of {\textquoteleft}preopened{\textquoteright} G{\textperiodcentered}C pairs with $H_2O_2$ and $H_2O$ molecules are significantly different, since the spatial structure of peroxide increases the parameter {\textquoteleft}propeller twist{\textquoteright}, taking the bases out of the plane of the pair.

It also should be noted that in the case of G{\textperiodcentered}C pair, there is still a configuration with the $H_2O_2$ molecule (Appendix, Fig. \ref{fig:Opened_H2O2}), which, on the one hand, is similar to the {\textquoteleft}preopened{\textquoteright} state because the parameter $R$ is almost unchanged, and the {\textquoteleft}opening{\textquoteright} parameter is significantly larger than in the Watson-Crick ones, but since the peroxide molecule is {\textquoteleft}embedded{\textquoteright} to the internal ($N_3 ... N_1$) hydrogen bond in this case, we call this state {\textquoteleft}opened{\textquoteright}. Note, that due to the geometry of the molecules, the {\textquoteleft}opened{\textquoteright} state occurs only for G{\textperiodcentered}C pair and only with $H_2O_2$ molecules.

\begin{table*}[t]
\begin{center}

\noindent\caption{The values of the distances $\Delta R= R-R_{WC}$ ($R$ - distance (in {\AA}) between $C_1'$ atoms in the corresponding pair (Fig. \ref{fig:WC}), $R_{WC}$ - the same distance in WC pair); parameters {\textquoteleft}opening{\textquoteright} and {\textquoteleft}propeller twist{\textquoteright} (in $degrees$), as well as the interaction energies E (in $kcal / mol$) for {\textquoteleft}closed{\textquoteright}, {\textquoteleft}preopened{\textquoteright}, {\textquoteleft}opened{\textquoteright} and {\textquoteleft}stretched{\textquoteright} configurations of base pairs with $H_2O_2$ and $H_2O$ molecules.$^{*}$\tnote{1}}\vskip3mm\tabcolsep4.5pt
\begin{threeparttable}[t]
\noindent{\footnotesize
\label{tab:results}
\begin{tabular}{ccccccccccccccc|}
 \hline%
State & & & & & & $\Delta R$  & & {\textquoteleft}opening{\textquoteright} & & {\textquoteleft}propeller twist{\textquoteright} & & $E$ \\
  \hline%
  \multirow{7}{*}{{\textquoteleft}\underline{closed}{\textquoteright} } & &  & &  & &  & &  & &  & &    \\
  & & \multirow{2}{*}{A{\textperiodcentered}T} & & $H_2O$ & & $0.1$ & & $-5.6$ & & $-2.6$ & & $-14.7$ \\
  & &  & & $H_2O_2$ & & $0.1$ & & $-5.6$ & & $-2.7$ & & $-16.0$ \\

  & &  & &  & &  & &  & &  & &  \\
  & &  \multirow{2}{*}{G{\textperiodcentered}C} & & $H_2O$ & & $0.0$ & & $-1.7$ & & $-0.5$ & & $-21.2$  \\

  & &  & & $H_2O_2$ & & $0.0$ & & $-1.7$ & & $0.0$ & & $-22.1$ \\
     & &  & &  & &  & &  & &  & &     \\
    \hline
    \multirow{7}{*}{{\textquoteleft}\underline{preopened}{\textquoteright}} & &  & &  & &  & &  & &  & &   \\
    & & \multirow{2}{*}{A{\textperiodcentered}T} & & $H_2O$ & & $-1.7$ & & $36.0$ & & $3.7$ & & $-15.8$  \\
    & &  & & $H_2O_2$ & & $-1.6$ & & $35.0$ & & $12.9$ & & $-16.6$ \\

    & &  & &  & &  & &  & &  & & \\
    & & \multirow{2}{*}{G{\textperiodcentered}C} & & $H_2O$ & & $-0.3$ & & $23.3$ & & $-0.8$ & & $-16.9$ \\

   & &  & & $H_2O_2$ & & $-0.5$ & & $17.2$ & & $42.9$ & & $-18.6$  \\
   & &  & &  & &  & &  & &  & &   \\
  \hline
   & &  & &  & &  & &  & &  & &  \\
  {\textquoteleft}\underline{opened}{\textquoteright} & & G{\textperiodcentered}C & & $H_2O_2$ & & $-0.1$ & & $47.0$ & & $7.3$ & & $-19.0$ \\
     & &  & &  & &  & &  & &  & &  \\
  \hline
  \multirow{7}{*}{{\textquoteleft}\underline{stretched}{\textquoteright}} & &  & &  & &  & &  & &  & & \\
     & & \multirow{2}{*}{A{\textperiodcentered}T} & & $H_2O$ & & $3.4$ & & $-35.9$ & & $1.8$ & & $-10.9$  \\

  & &  & & $H_2O_2$ & & $3.1$ & & $-11.7$ & & $-69.1$ & & $-17.6$  \\

   & &  & &  & &  & &  & &  & & \\
  & & \multirow{2}{*}{G{\textperiodcentered}C} & & $H_2O$ & & $3.1$ & & $-31.5$ & & $0.0$ & & $-13.0$ \\

  & &  & & $H_2O_2$ & & $2.9$ & & $-7.0$ & & $-72.4$ & & $-20.9$ \\
  & &  & &  & &  & &  & &  & & \\
  \hline

\end{tabular}
\begin{tablenotes}\footnotesize
${^*}$ All the values are rounded to the first decimal due to the accuracy of the parameters that are used for the calculations of the corresponding structures.
\end{tablenotes}
}
\end{threeparttable}
\end{center}
\end{table*}

Configurations of {\textquoteleft}stretched{\textquoteright} pairs with water and hydrogen peroxide differ significantly by their parameters (Appendix, Fig. \ref{fig:Stretched_H2O_H2O2}).
It should be noted that for both the A{\textperiodcentered}T and the G{\textperiodcentered}C base pairs, the interaction energy of the {\textquoteleft}stretched{\textquoteright} pair for the
complex with water molecule is lower than the corresponding value for the {\textquoteleft}preopened{\textquoteright} configuration, and for complexes with hydrogen
peroxide molecule, on the contrary, it is substantially higher (Tabl. \ref{tab:results}).

\section{About the possibility of an experimental observation of the formation of complexes of hydrogen peroxide molecules with the DNA base pairs}
\label{experiment}

During last decades the technique that allows to study features of single molecules was improved significantly. With the
help of single-molecule micromanipulation methods important properties of a DNA macromolecule, such as stretching,
bending, twisting~\cite{bustamante2003}, and the consequent opening of nucleic base pairs under the action of external force (unzipping)~\cite{bockelmann2002,ritort2010} can be investigated. The experiment is carried out at a constant opening velocity, and the dependence of
the opening force on the displacement value is measured. At the beginning, the force is not enough to open a double
helix, so it sharply increases, and then reaches the plateau, which corresponds to the beginning of the opening of the
base pairs.

In the work~\cite{ourWorkUnz}, it has been shown that during the unzipping of a double helix, depending on the opening
velocity, base pairs can open as along the {\textquoteleft}stretch{\textquoteright} pathway, as well as along the {\textquoteleft}opening{\textquoteright} pathway. And in the work~\cite{volkovMolDynSym} it was shown that during unzipping process configurations of base pairs that are stabilized by water bridges occur.  In this regard,
if a certain concentration of hydrogen peroxide is added to the solution, {\textquoteleft}preopened{\textquoteright}, and {\textquoteleft}stretched{\textquoteright} pairs stabilized
by $H_2O_2$ molecules can appear in the conditions of this experiment. As it follows from our calculations, the binding energy of these complexes is significantly higher than the energy of the same complexes with the water molecule.
Therefore, the opening force of a double helix in the experiment carried out in the presence of a certain concentration of $H_2O_2$ molecules  should be higher as a result of the interaction of the
base pairs with $H_2O_2$ molecules. The observation of these force growing in the unzipping experiment can serve as the proof of our hypothesis about the blocking of the DNA base pairs by hydrogen peroxide molecules.

Also the complexes of hydrogen peroxide with DNA base pairs can be observed by the methods of Raman spectroscopy. It is known that all DNA atomic groups have their own vibrational frequency. As known, the vibration frequencies of atomic groups of DNA nucleic bases are in the range of $\sim 1500$ $cm^{-1}$~\cite{prescott1984DNAvibrationSpectra}. Interaction of base pairs with water molecules can slightly lower frequency and amplitude of the vibrations~\cite{maleev1993}. Therefore, the interaction of hydrogen peroxide molecules with DNA base pairs must manifest itself as the shifting of the absorption peak to the less-frequency range and lower its height with comparison to the same complexes with water molecules.

\section{Discussion and conclusions}

\begin{figure}[h!]% figure* for wide figure, [h] [!] to change the placement
\centering
\vskip1mm
\includegraphics[width=10cm]{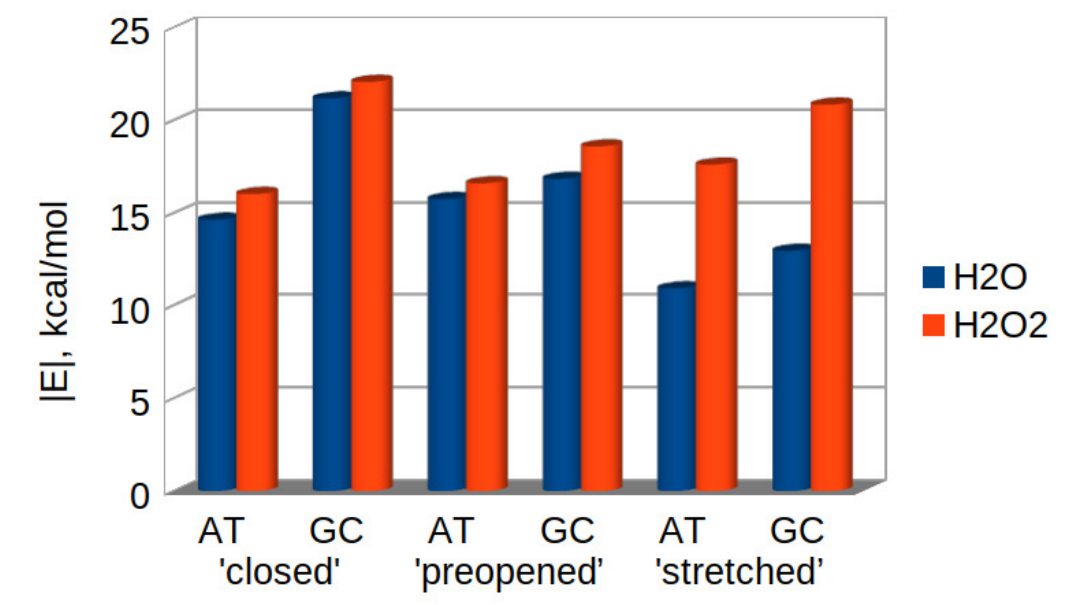}
\vskip-3mm\caption{Diagram of the interaction energy values for complexes consisting of hydrogen peroxide and water molecules with {\textquoteleft}closed{\textquoteright}, {\textquoteleft}preopened{\textquoteright} and {\textquoteleft}stretched{\textquoteright} configurations of A{\textperiodcentered}T and G{\textperiodcentered}C base pairs.}
\label{fig:diagram}
\end{figure}

In the present paper, the spatial configurations of complexes consisting of nucleic base pairs with hydrogen peroxide
and water molecules are investigated. The comparison of the interaction energy values is schematically shown on
the diagram (Fig. \ref{fig:diagram}). As it can be seen from the results, there are  configurations of A{\textperiodcentered}T and G{\textperiodcentered}C base pairs ({\textquoteleft}preopened{\textquoteright} and {\textquoteleft}stretched{\textquoteright}), which are stabilized by hydrogen peroxide molecules much better than by water molecules.
 
Interaction of hydrogen peroxide molecules with DNA base pairs can manifest itself in living cell. Namely, in the process of DNA replication, when two DNA macromolecules are formed from one double helix. At the initial stage of this process, an enzyme passes along a double helix~\cite{galletto2004} and consequently opens its base pairs one after another. DNA bases interact  with water molecules all the time. But, as it was mentioned above, significant amount of $H_2O_2$ molecules are introduced to the medium during ion beam therapy. If the interaction energy of nucleic bases of the pair with hydrogen peroxide molecule is large enough compared to the same interaction energy with water molecule, the DNA unzipping by enzyme can be terminated.

As can be seen from the obtained results, the most significant difference in interaction energies is for {\textquoteleft}stretched{\textquoteright} configurations of the pairs. Therefore, the possibility of blocking {\textquoteleft}stretched{\textquoteright} pairs by hydrogen peroxide molecules is significantly more probable. In this case the difference between the opening energies is ${\approx} 7-8$ $kcal/mol$ (Tabl. \ref{tab:results}). This is due to the fact that the hydrogen peroxide molecule, because of its spatial structure, forms four hydrogen bonds with nucleic bases (Appendix, Fig. \ref{fig:Stretched_H2O_H2O2} b,d). At the same time, water molecule forms three hydrogen bonds, and two of them are substantially curved, that is, their energy is weakened (Appendix, Figures \ref{fig:Stretched_H2O_H2O2} a,c). It should also be noted that, since in {\textquoteleft}stretched{\textquoteright} configurations the parameter {\textquoteleft}propeller twist{\textquoteright} is significant, the formation of such configurations is possible only in the unzipping fork, where there is no stacking interaction from one side of the pair.

It should be noted, that the energy values obtained in the present work are only the enthalpy values, entropy is not taken into account.  However, due to the similar structure of $H_2O_2$ and $H_2O$ molecules, the entropy contribution to the interaction of this molecules with nucleic bases, should also be similar. Therefore, our approach allows us to obtain a qualitative picture of the formation of complexes of
hydrogen peroxide molecules with base pairs, and to see a essential energy advantage compared to the same complexes with a water molecule. 

Formation of complexes of $H_2O_2$ molecules with DNA  can completely block the DNA transcription of cancer cells, and can be a key factor of the action of high-energy ions on cancerous tumors in the process of ion beam therapy.

\vskip3mm \textit{Acknowledgement.}
The present work was partially supported by the Program of Fundamental Research of the Department of Physics and Astronomy of the National Academy of Sciences of Ukraine (project number 0116U003192).

\section*{APPENDIX}

\setcounter{figure}{0}
\makeatletter
\renewcommand{\thefigure}{A\@arabic\c@figure}
\makeatother

\begin{figure}[h!]% figure* for wide figure, [h] [!] to change the placement
\begin{center}
\vskip1mm
\includegraphics[width=7cm]{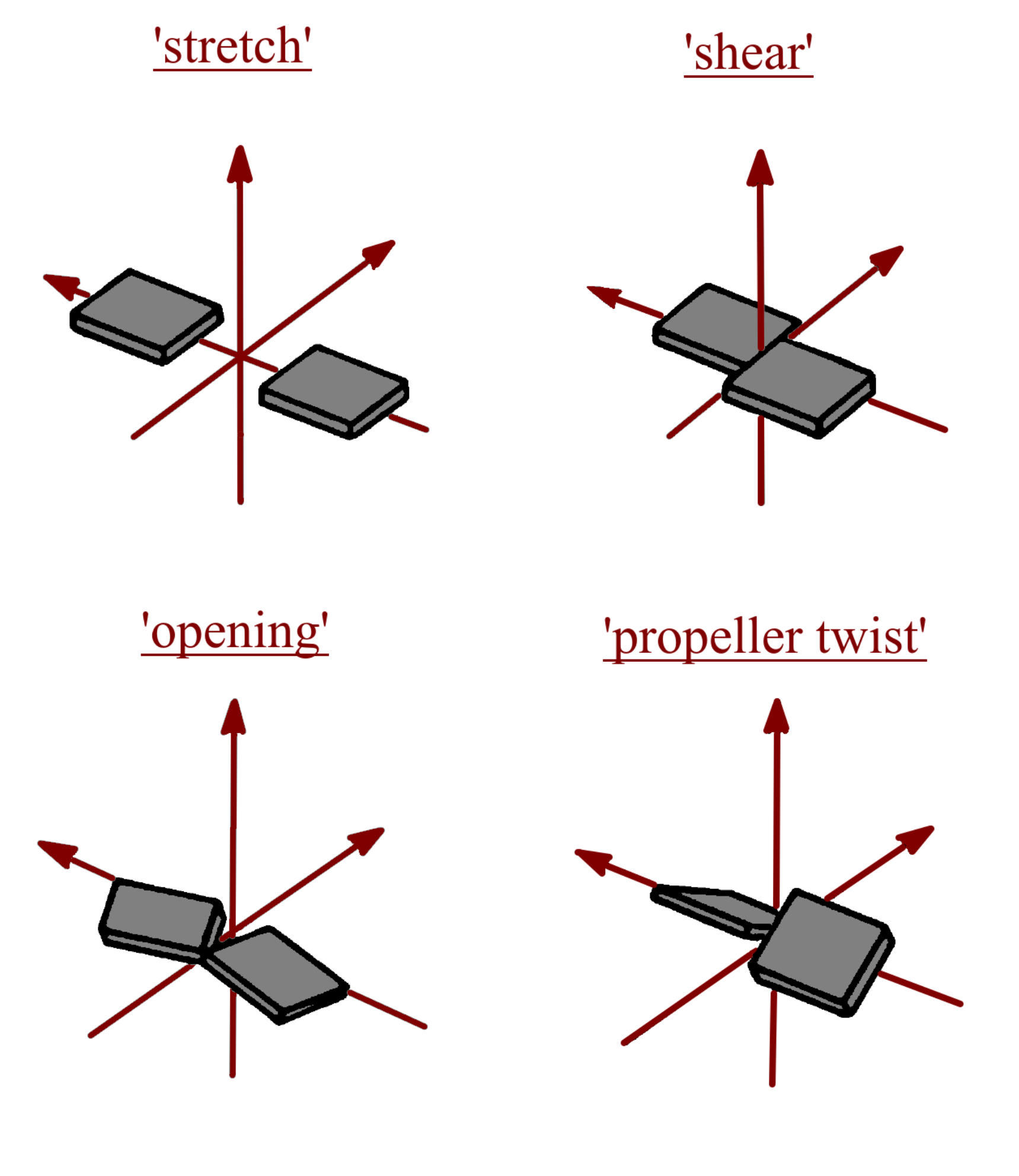}
\vskip-3mm\caption{Degrees of freedom of nucleic bases in a complementary pair, which were taken into account in the calculations of molecular complexes in this paper.}
\label{fig:pathways}
\end{center}
\end{figure}

\begin{figure}[h!]% figure* for wide figure, [h] [!] to change the placement
\vskip1mm
\begin{center}
\includegraphics[width=9cm]{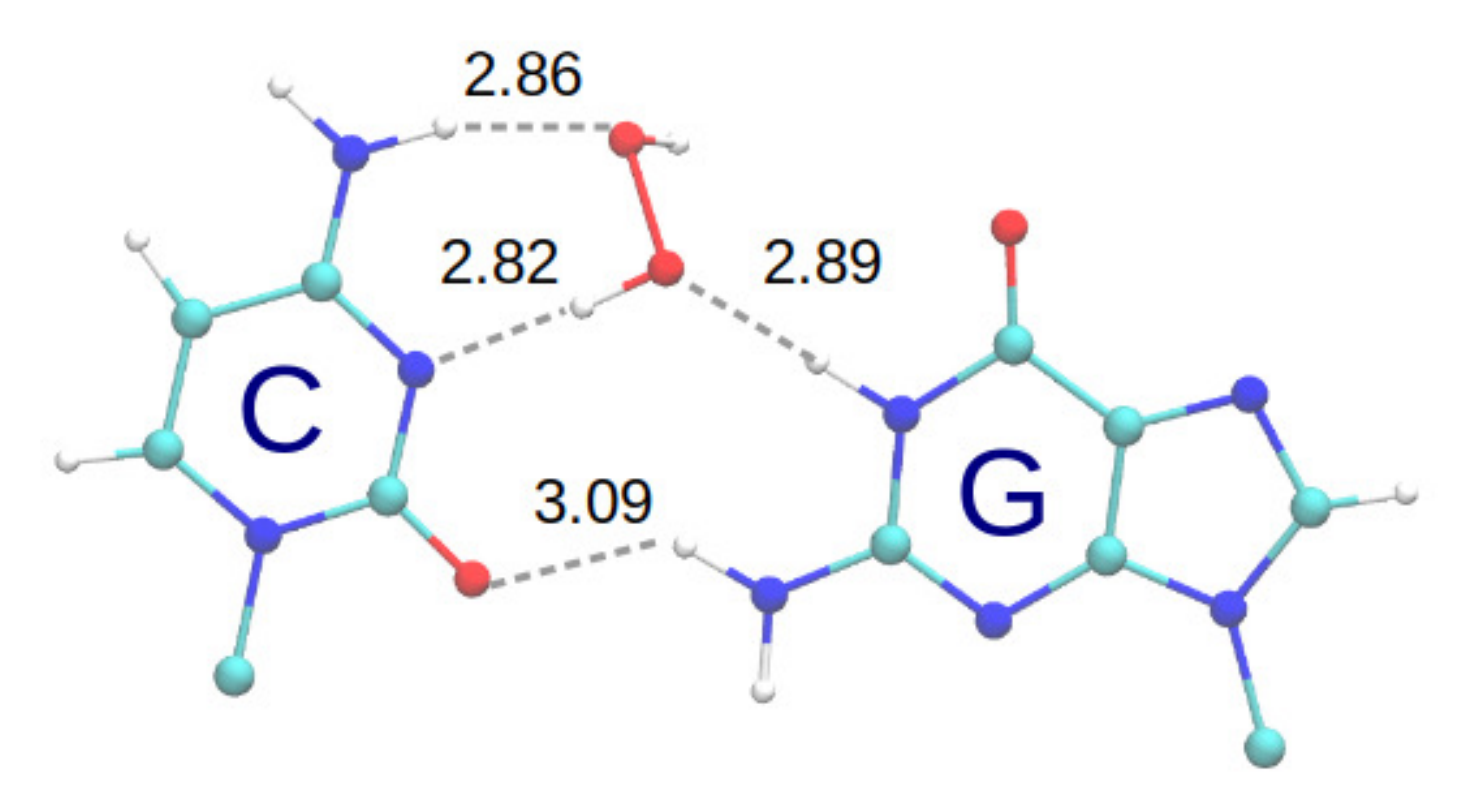}
\vskip-3mm\caption{{\textquoteleft}Opened{\textquoteright} configuration of G{\textperiodcentered}C pair with $H_2O_2$ molecule calculated in the present work. The numbers indicate the distance (in \AA) between heavy atoms in the corresponding hydrogen bonds.}
\label{fig:Opened_H2O2}
\end{center}
\end{figure}

\begin{figure}[h!]% figure* for wide figure, [h] [!] to change the placement
\vskip1mm
\includegraphics[width=17cm]{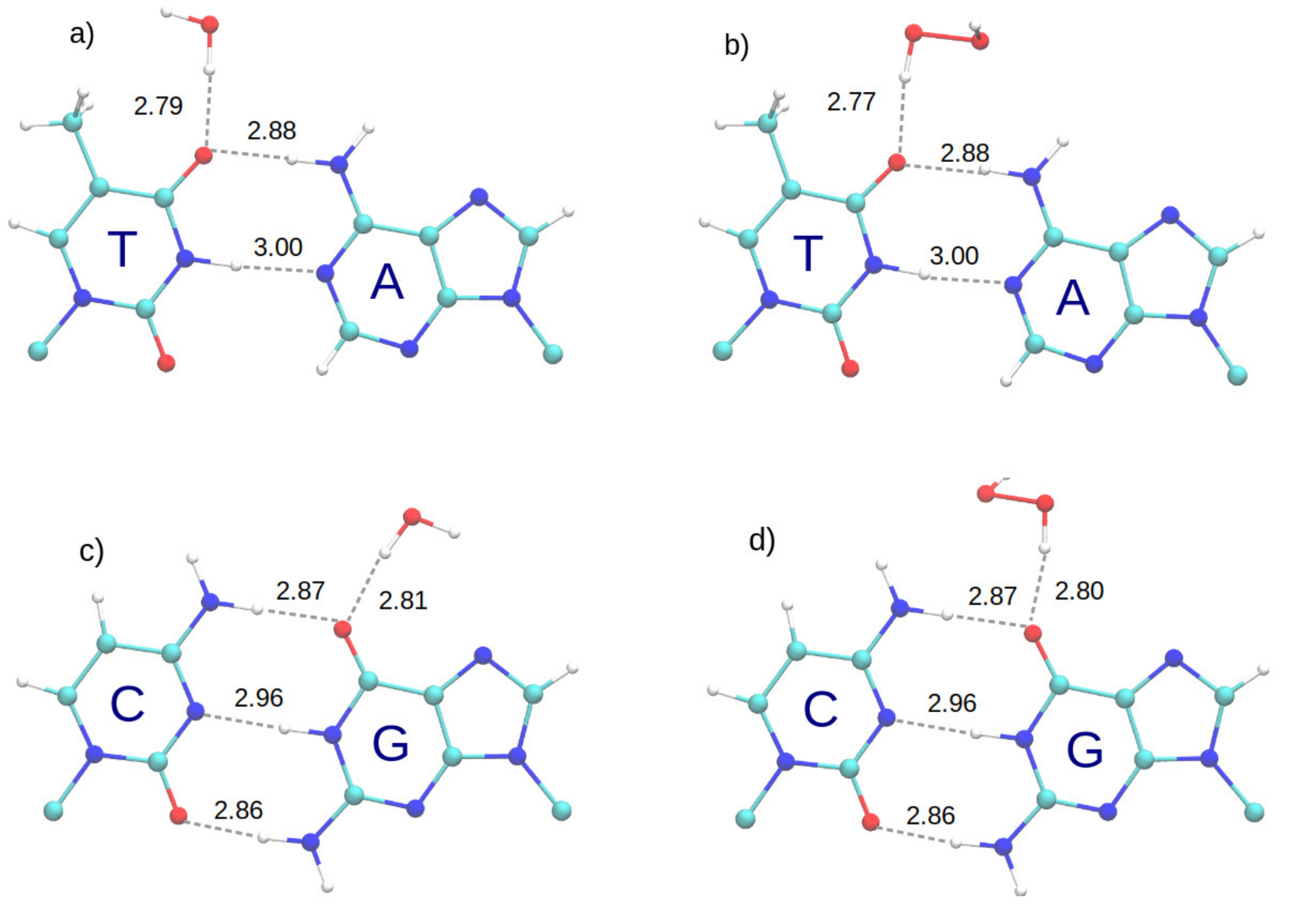}
\vskip-3mm\caption{ Complexes of complementary pairs of A{\textperiodcentered}T and G{\textperiodcentered}C with water and hydrogen peroxide molecules, bound from the major groove ({\textquoteleft}closed{\textquoteright} pair): a) A{\textperiodcentered}T with $H_2O$ molecule; b) A{\textperiodcentered}T with $H_2O_2$; c) G{\textperiodcentered}C with $H_2O$ d) G{\textperiodcentered}C with $H_2O_2$. }
\label{fig:Closed_H2O_H2O2}
\end{figure}

\begin{figure}[h!]% figure* for wide figure, [h] [!] to change the placement
\vskip1mm
\includegraphics[width=17cm]{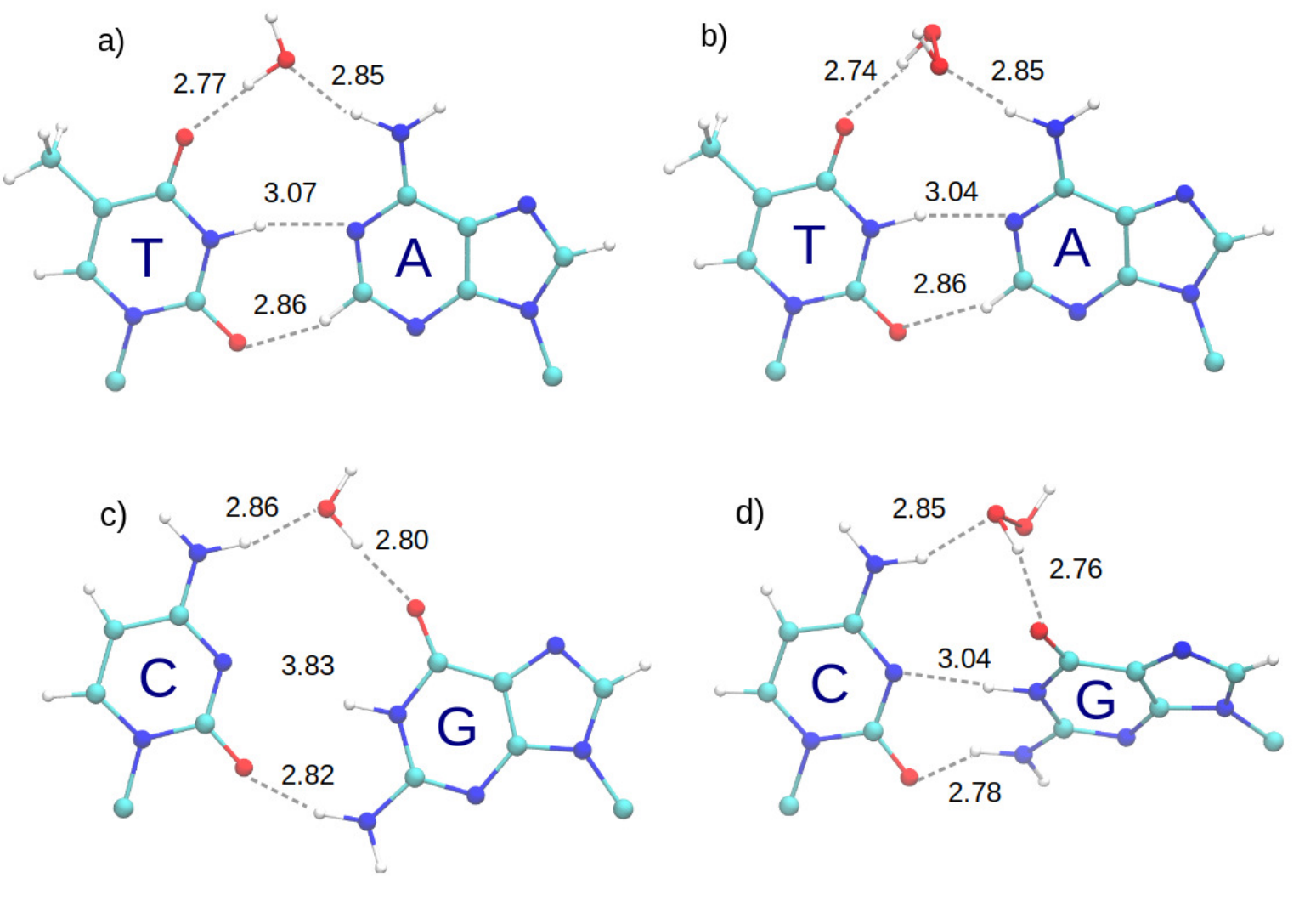}
\vskip-3mm\caption{The stable configurations of the {\textquoteleft}preopened{\textquoteright} base pairs calculated in the present work: A{\textperiodcentered}T with water (a) and hydrogen peroxide (c) molecules; G{\textperiodcentered}C with water (c) and hydrogen peroxide (d) molecules. The numbers indicate the distance (in \AA) between heavy atoms in the corresponding hydrogen bonds.}
\label{fig:Preopened_H2O_H2O2}
\end{figure}

\begin{figure}[h!]% figure* for wide figure, [h] [!] to change the placement
\centering
\vskip1mm
\includegraphics[width=18cm]{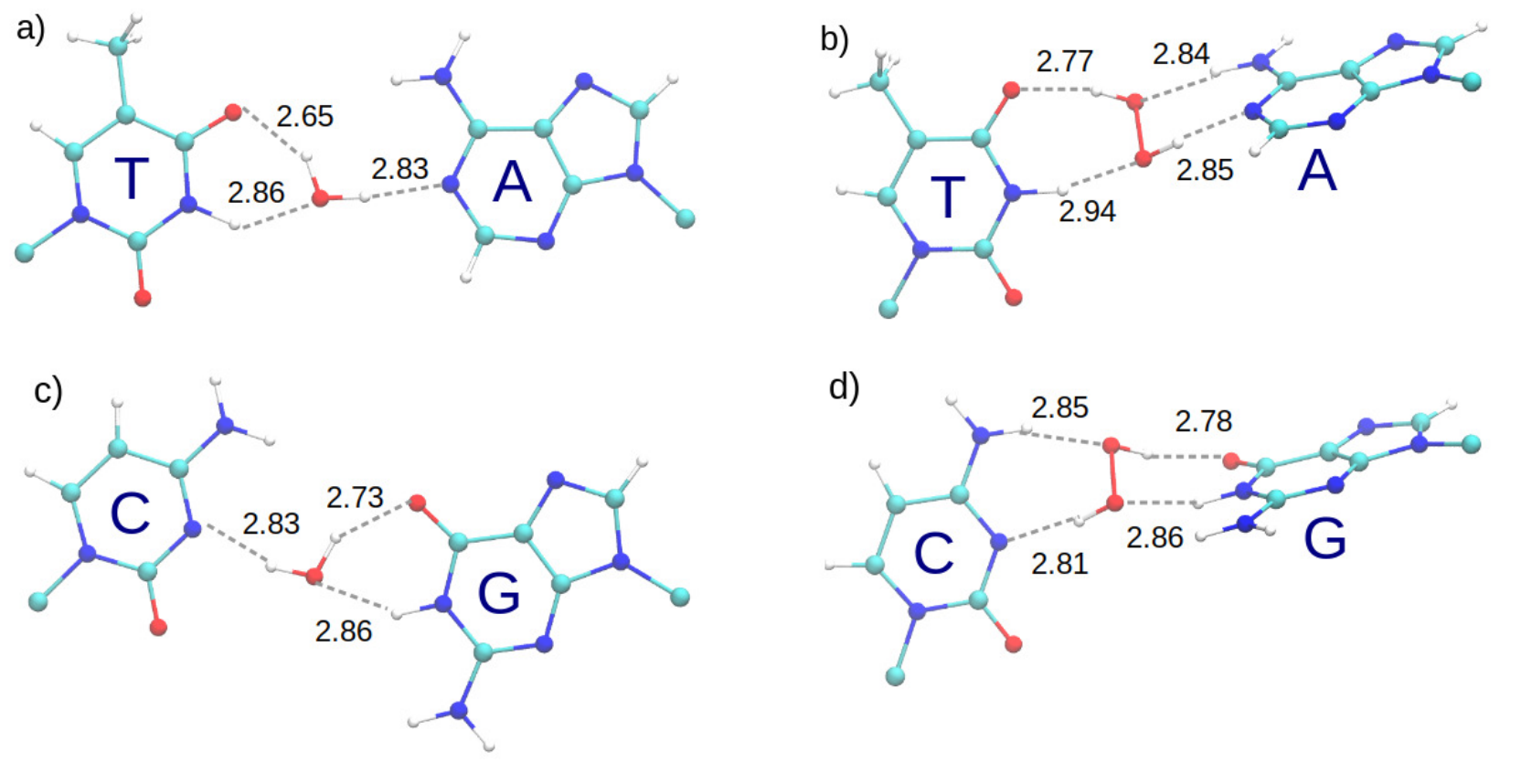}
\vskip-3mm\caption{The stable configurations of {\textquoteleft}stretched{\textquoteright} base pairs calculated in the present work: A{\textperiodcentered}T with water (a) and hydrogen peroxide (c) molecules; G{\textperiodcentered}C with water (c) and hydrogen peroxide (d) molecules. The numbers indicate the distance (in \AA) between heavy atoms in the corresponding hydrogen bonds.}
\label{fig:Stretched_H2O_H2O2}
\end{figure}

\end{document}